# Predicting Personality from Book Preferences with User-Generated Content Labels

Ng Annalyn, Maarten W. Bos, Leonid Sigal, and Boyang Li

**Abstract**—Psychological studies have shown that personality traits are associated with book preferences. However, past findings are based on questionnaires focusing on conventional book genres and are unrepresentative of niche content. For a more comprehensive measure of book content, this study harnesses a massive archive of content labels, also known as 'tags', created by users of an online book catalogue, *Goodreads.com*. Combined with data on preferences and personality scores collected from *Facebook* users, the tag labels achieve high accuracy in personality prediction by psychological standards. We also group tags into broader genres, to check their validity against past findings. Our results are robust across both tag and genre levels of analyses, and consistent with existing literature. Moreover, user-generated tag labels reveal unexpected insights, such as cultural differences, book reading behaviors, and other non-content factors affecting preferences. To our knowledge, this is currently the largest study that explores the relationship between personality and book content preferences.

**Index Terms**—Personality Profiling, Narrative Preferences, Social Media, Behavioural Footprints

——————————— ◆ ———————————

## 1 INTRODUCTION

*"Histories make men wise;*

*poets witty;*

*the mathematics subtle;*

*natural philosophy deep;*

*moral grave;*

*logic and rhetoric able to contend."*

*By Francis Bacon, Of Studies (1597)*

Francis Bacon may have been the first to suggest a correlation–perhaps even a causal relation–between book preferences and the personality of readers. Indeed, research has found that reading fiction leads to changes in personality [1] and increased empathy [2]. While book reading may influence personality, personality in turn may affect book choice. This is supported by correlations found between personality and book preferences [3]. Being able to predict book preferences using readers' personality has many potential applications, such as personalizing products and services, improving recommender systems, and enabling targeted advertising.

However, due to difficulties in data collection, research on personality and book preferences typically focus on a few dozen book genres or less, such as having four genres for novels [4], 16 genres for books [5], or 34 genres for books and magazines combined [6]. The largest study [3] to our knowledge inspected 81 book topics and their correlations with readers' personality. Narrow categorizations of book content can be problematic as preferences for niche genres may be inaccurately inferred. Moreover, studies measured book preferences using self-report questionnaires, which can be lengthy and thus vulnerable to errant or null responses [7].

To improve both the quantity and quality of data for our study, we combine two online data sources. For data on book content, we use over 24,000 user-supplied tags from a book catalogue website, *GoodReads.com*. For data on reader personality and book preferences, we use a database of *Facebook* profiles comprising more than 60,000 respondents who had 'liked' book-themed *Facebook* pages and who had also completed a personality survey on the social networking site [8].

We adopt the Big Five personality model, also known as the five-factor model, which consists of extraversion, agreeableness, openness, neuroticism and conscientiousness. This set of five traits is known to predict a wide range of behaviors and psychopathology [9]. We briefly review their known associations with book preferences here:

**Extraverts** enjoy social activities and have high arousal levels [9]. They prefer content related to social activities such as parties [10], as well as arousing content such as horror [6]. Hence, we hypothesize that extraversion would be associated with a preference for genres with socially oriented themes, as well as genres which are stimulating.

**Agreeable** individuals are kind and considerate [9], and tend to empathize with story characters [11]. They prefer narratives on positive social relationships, such as romance and family [6], hence they are likely to steer clear of violent

————————————————
- N. Annalyn is with the Ministry of Defence (Singapore), #16-01 Defence Technology Tower B, Singapore 109681.
  E-mail: ng_li_ting_annalyn@defence.gov.sg.
- M. W. Bos, L. Sigal, and B. Li are with Disney Research Pittsburgh, 4720 Forbes Avenue, Lower Level, Suite 110, Pittsburgh, PA 15213.
  E-mail: {mbos, lsigal, albert.li }@disneyresearch.com.

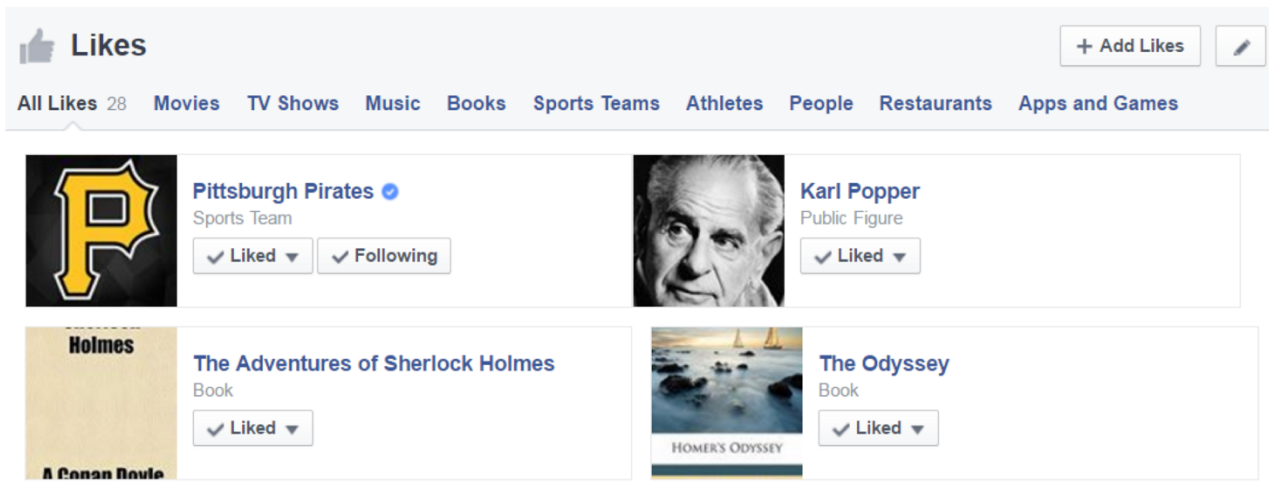

Fig. 1. Facebook 'likes' for various pages.

or disturbing themes. As agreeable people tend to evaluate media content favorably in general [10], we might also expect them to 'like' more books on average.

**Open** individuals seek intellectual stimulation and are comfortable with new ideas [9]. Openness predicts a preference for avant-garde genres [12], and has been a consistent predictor of fiction exposure [2], [3], [4], [13]. Based on existing literature, we hypothesize that individuals with high openness would appreciate intellectually stimulating and fiction genres. Since reading is an intellectual activity, we would also expect open individuals to 'like' more books in general.

**Neurotic** individuals are emotionally unstable. Being prone to feeling lonely and depressed [14], they may use media narratives as a means of escape from everyday life [4]. Hence, we hypothesize that neurotic individuals would prefer narratives about feel-good, alternative realities.

**Conscientious** individuals are achievement-striving and self-disciplined, preferring deliberate planning over spontaneity [9]. They have also been reported to like non-fiction content such as news and politics [6], [15]. Hence, we hypothesize that conscientiousness would be associated with a preference for books of informative and practical content.

Merging personality data (from Facebook surveys) with book data (from Facebook 'likes' and GoodReads), we conduct three levels of analysis. The first is a tag-level analysis, in which we correlate personality with book tags. As tags were spontaneously generated by readers themselves, they contain richer information on book content compared to the usual, smaller set of genre categories. Second is a genre-level analysis. To verify that our findings are consistent with those from previous studies, tags are clustered into broader book genres and then correlated with personality again. Third, we examine whether personality has an overarching influence on one's tendency to like books in general.

## 2 DATA COLLECTION

### 2.1 Personality and Book Preferences

We use data collected from a Facebook app called *myPersonality* [8], which allows users to measure their Big Five personality traits with the International Personality Item Pool questionnaire [16]. Because users received feedback on their personality scores, they were likely to be motivated to respond diligently.

Besides their personality scores, users of the *myPersonality* app also shared which Facebook pages they had 'liked'. Facebook pages can be dedicated to any entity, such as a book, movie, or celebrity (see Fig. 1 for an example). For our study, we focus on pages labeled as books. This enables us to examine the correlation between users' personality scores and their book preferences.

Data collected via Facebook has been shown to be comparable to data collected via standalone websites [17]. Moreover, the Facebook personality dataset we use in this study has successfully predicted a range of personal traits, from web browsing habits [18] to language use [19]. Youyou et al. even suggested that personality inferences of users made based on digital footprints such as Facebook 'likes' are more accurate than those made by users' friends [20]. Due to its wide adoption, we deem the reliability of this dataset to be satisfactory for our study.

We use books pages with at least 50 'likes' from Facebook users who also completed the personality questionnaire. This achieves more reliable personality profile estimates of people who liked each book. In all, we analyze 479 books that were 'liked' by 61,662 users. For each of the five personality dimensions, we took the median score of all users who 'liked' the book as the aggregate personality score for that book. Median scores were favored over mean scores to reduce the influence of outliers.

### 2.2 Book Content

To extract a book's content, we adopt a data-driven approach by mining user-generated tags from GoodReads, an online book catalogue. When this study was conducted in 2016, the site had more than 40 million users and more than 1.3 billion books. Users can label books with descriptive tags, which cover a broad range of concepts like genre (e.g. `children's-literature`), time of publication (e.g. `20th-century-fiction`), story characters (e.g. `Dumbledore`), author information (e.g. `British-author`), awards (e.g. `orange-prize`) and reading behavior (e.g., `back-burner`). Due to their richness, we chose these user-generated tags as a proxy for book content. However, some tags, such as `upstairs-bookshelves`, appear to make sense only to a small group of users. This calls for robust analysis techniques that can withstand noise.

With permission from Goodreads, we crawl their site for the 479 books in our personality dataset, and then harvest the tags which Goodreads users had associated with these books. Besides accounting for the books present in the personality dataset, we also identify the top 50 books associated with each of the top 2000 most frequently-used tags across the whole catalogue. We then crawl all tags associated with these books.

Next, we match book titles from Goodreads to their respective pages on Facebook, leading to a many-to-many relationship. For example, the book *Harry Potter and the Philosopher's Stone* is matched not only to the Facebook page of the same name, but also to a general page for the *Harry Potter* series. At the same time, the general Facebook page for *Harry Potter* is matched to all seven books in the series.

## 3 DATA PROCESSING

Goodreads users can create their own tags. While this provides a rich source of information, it also introduces noise that poses several challenges for analysis. To overcome these challenges, we employ several techniques.

First, we use a set of criteria to filter tags for analysis:

- For each book crawled, only tags that applied 3 times or more are recorded.
- Tags applied less than 50 times in total and tags applied to less than 15 books are discarded.
- Tags must consist of at least 3 characters, at least 1 letter, and at most 2 non-English characters.

We use these filtering criteria because they were deemed via manual inspection to be effective at eliminating non-informative tags. After crawling and filtering, our dataset contained 14,731 unique books, 24,091 unique tags and 193,498,469 total tags.

Next, we identify four challenges in analyzing the tags data:
- **Information Value**. Common tags (e.g. `fiction`, `book-club`, and `favorites`) appear frequently across many titles, and thus are not useful in distinguishing between books.
- **Synonyms**. Some tags have identical or similar meanings (e.g. `children` and `kids`), and hence need to be analysed as one.
- **Idiosyncrasies**. Some tags are used whimsically. For example, *Harry Potter and the Philosopher's Stone* was tagged as `science` more than 20 times.
- **Random Noise**. We expect a baseline level of random noise. If a tag is applied to a book 20 times, and to another for 21 times, this difference would likely be due to random fluctuations rather than actual difference in content.

To distinguish informative tags, we use the term frequency-inverse document frequency (tf-idf) measure. With tf-idf, the frequency $f_{t,b}$ of a tag appearing in a book is discounted by how common the tag $t$ generally is. In other words, common tags such as `fiction` and `favourites` are discounted heavily to indicate their low information value. Letting $f_{t,b}$ denote the frequency of tag $t$ appearing in book $b$, $N$ denote the total number of books, $n_t$ denote the number of unique books that tag $t$ is applied to, and we have

$$\text{tf-idf}(t, b) = f_{t,b} \log\left(1 + \frac{N}{n_t}\right) \quad (1)$$

Using tf-idf, we can build a book-by-tag matrix, $M$. In $M$, each row represents a book, each column represents a tag, and each entry represents the corresponding tf-idf value.

Next, we group similar tags together. We do this by combining results from two similarity measures.

The first similarity measure is derived from the co-occurrence of tags in books. That is, if two tags occur in similar books, the tags are likely to share similar meanings and belong to the same genre. We compute a low-rank approximation of $M$, matrix $\widehat{M}$. Formally, we minimize the following objective:

$$\widehat{M}^* = \min_{\widehat{M}} \left\|M - \widehat{M}\right\|_F \text{ s.t. rank}(\widehat{M}) \leq \lambda \quad (2)$$

where the $\lambda$ is the desired rank of $\widehat{M}$ and $\|\cdot\|_F$ is the Frobenius norm. The minimization is achieved using singular value decomposition. Each tag $t$ is represented as a column vector $\boldsymbol{v}_t$ in $\widehat{M}$. The similarity between two tags $t$ and $t'$ is then computed as the cosine of the angle in between:

$$\text{similarity}(t, t') = \frac{\boldsymbol{v}_t \cdot \boldsymbol{v}_{t'}}{\|\boldsymbol{v}_t\|\|\boldsymbol{v}_{t'}\|} \quad (3)$$

Although the above captures co-occurrence between tags, we also want to directly capture lexical similarity. Thus, we derive a second similarity measure based on shared words between tags (e.g. between `historical-novel` and `historical-fiction`). Each word in a tag is first lemmatized using ClearNLP [21]. As in co-occurrence similarity, we compute a tag-by-word matrix using tf-idf to discount frequent words, followed by a low-rank approximation of the matrix. Similarity between tags can be computed as the cosine distance between row vectors in this matrix.

Overall similarity is computed as a weighted sum of co-occurrence-based (95% weight) and word-lemma-based (5% weight) similarities. Then, we use the OPTICS clustering algorithm [22] to cluster similar tags together and to discard tags that do not fit into any cluster. A round of manual coding is performed to correct any errors in the clustering, resulting in a total of 396 tag clusters, where each tag cluster corresponds to a single semantic meaning. Each tag cluster is then labelled with a semantically representative tag as a label for book content, and henceforth

TABLE 1
TAG CLUSTERS MOST CORRELATED WITH PERSONALITY

| | r | | | | | | r | |
|---|---|---|---|---|---|---|---|---|
| *** | -0.17 | horror-and-gothic | Introverted | Extraversion | Extraverted | relationships | 0.25 | *** |
| ** | -0.16 | fantasy-sci-fiction | | | | chick-lits | 0.17 | *** |
| ** | -0.15 | parallel-world | | | | memoir-autobio | 0.17 | *** |
| ** | -0.14 | other-supernatural | | | | celebrity-romance | 0.14 | ** |
| ** | -0.13 | manga-collection | | | | african-american-lit | 0.13 | ** |
| *** | -0.17 | italian-renaissance | Disagreeable | Agreeableness | Agreeable | christian-classics | 0.24 | *** |
| ** | -0.15 | cult-classics | | | | relationships | 0.20 | *** |
| ** | -0.15 | psychological-drama | | | | family-drama | 0.15 | ** |
| * | -0.12 | scary-stuff | | | | kids-book | 0.13 | ** |
| * | -0.11 | japanese-culture | | | | buddhism | 0.12 | * |
| *** | -0.26 | light-fantasy | Traditional | Openness | Open | back-burner | 0.28 | *** |
| *** | -0.19 | grade-4-6 | | | | philosophical-novel | 0.25 | *** |
| *** | -0.17 | indian-books | | | | university-readings | 0.24 | *** |
| *** | -0.17 | chick-lits | | | | classic-favs | 0.19 | *** |
| * | -0.11 | christian-classics | | | | plays-and-musicals | 0.14 | ** |
| *** | -0.24 | theology-religion | Levelheaded | Neuroticism | Neurotic | mental-issues | 0.25 | *** |
| *** | -0.22 | politics-and-philosophy | | | | pretty-writing | 0.24 | *** |
| *** | -0.20 | professional-reading | | | | sad-endings | 0.20 | *** |
| *** | -0.19 | leadership-business | | | | paranormal-fantasy-scifi | 0.17 | *** |
| *** | -0.16 | science-and-technology | | | | dark-and-dangerous | 0.15 | ** |
| ** | -0.16 | modern-fantasy | Tardy | Conscientiousness | Conscientious | grown-up-stuff | 0.23 | *** |
| ** | -0.13 | graphica | | | | brain-food | 0.23 | *** |
| ** | -0.13 | teenage-books | | | | history-ww2 | 0.21 | *** |
| * | -0.12 | fantasy-sci-fiction | | | | professional-reading | 0.19 | *** |
| * | -0.11 | humor-comedy | | | | leadership-business | 0.18 | *** |

$r$ = correlation coefficient; * $p < 0.05$; ** $p < 0.01$; *** $p < 0.001$.

treated as a single tag for analysis against personality. Specifically, to consolidate tags belonging to the same cluster, the median of their tf-idf values is used.

As a single Facebook page can contain multiple books on Goodreads, we consolidate the book data by manually mapping Goodreads books to Facebook pages. For each book, we first normalize its feature vector comprising tf-idf values to unit length. Next, feature vectors of books referred to by the same Facebook page are summed; the resulting summed vector is then normalized to unit length again.

## 4 ANALYSIS

We conduct a two-level analysis to examine how personality predicts book content preferences at the tag level and at the genre level. We also analyze how personality could influence one's general tendency to like books.

### 4.1 Tag-Level Analysis

We compute correlations between the tf-idf values of each tag cluster and each of the Big Five personality dimensions.

Next, we perform lasso regression to predict personality from tag cluster features. Unlike regular regression, lasso regression maintains a higher prediction accuracy despite correlations between features (i.e. multi-collinearity) through regularization. For each personality trait, we use the regularization coefficient yielding the lowest mean squared test error from a 10-fold cross-validation.

We also perform the same prediction using a random forest regression with 500 trees. The technique involves simulating different combinations of features in multiple decision trees to select the best combination of features that predicts personality. As determined by cross validation, each tree utilizes 132 variables selected randomly. With the random forest regression, we compute the importance of each tag cluster feature based on the increase in mean squared prediction error when that feature is removed.

#### 4.1.1 Results

Table 1 shows the tag clusters that are most strongly correlated with each personality trait. All correlations shown are statistically significant at $p < 0.05$ and most are significant at $p < 0.001$. The most positive correlation is between the back-burner tag cluster and the openness trait ($r = 0.28$), while the most negative correlation is between the light-fantasy tag cluster and openness ($r = -0.26$).

Table 2 shows the tag clusters with the biggest absolute coefficients in the five lasso regression analyses predicting scores for each personality trait. Based on the $R^2$ values

TABLE 2
TOP LASSO REGRESSION COEFFICIENTS IN PERSONALITY PREDICTION

| Trait | Tag | β |
|---|---|---|
| Extraverted $R^2 = 0.32$ | relationships | 2.19 |
| | beach-reading | 1.16 |
| | south-africa | 0.77 |
| | manga-collection | -1.77 |
| | parallel-world | -2.11 |
| | forgotten_realms | -2.59 |
| Agreeable $R^2 = 0.34$ | christian-classics | 0.85 |
| | relationships | 0.84 |
| | family-drama | 0.56 |
| | dark-and-dangerous | -0.41 |
| | italian-renaissance | -0.91 |
| | psychological-drama | -0.99 |
| Open $R^2 = 0.41$ | philosophy-psychology | 0.66 |
| | university-readings | 0.53 |
| | philosophical-novel | 0.44 |
| | christian-classics | -0.52 |
| | girly-fiction | -0.52 |
| | light-fantasy | -1.32 |
| Neurotic $R^2 = 0.38$ | dark-and-dangerous | 1.41 |
| | mental-issues | 1.37 |
| | pretty-writing | 1.34 |
| | science-and-technology | -0.94 |
| | politics-economics | -1.36 |
| | professional-reading | -5.38 |
| Conscientious $R^2 = 0.30$ | professional-reading | 1.19 |
| | theology-religion | 0.84 |
| | brain-food | 0.77 |
| | modern-fantasy | -0.40 |
| | mental-issues | -0.55 |
| | graphica | -0.69 |

$R^2$ = coefficient of determination, or the proportion of variance explained by the lasso regression model; β = regression coefficient.

from each analysis, book content seems best at predicting scores on the openness trait.

Fig. 2 shows the tag clusters that result in the largest decreases in mean squared error in the five random forest regression analyses predicting scores for each personality trait. Green and red colors represent positive and negative correlations respectively between clusters and traits. Note that most of the top predictive tag clusters for agreeableness have positive correlations with the trait.

Results from correlation, lasso regression and random forest regression analyses are largely consistent. For example, `fantasy-sci-fiction` has a strong negative correlation with extraversion (Table 1), and this is supported by the lasso regression predicting extraversion, which shows strong negative coefficients for fantasy settings such as `parallel-world` and `forgotten_realms`, (Table 2). This is again supported by random forest regression findings that the second most important variable in predicting extraversion is `fantasy-sci-fiction` (Fig. 2). Differences between Tables 1 and 2 may be attributed to the use of L1-regularization in lasso. The regularization penalizes the number of non-zero coefficients, forcing the algorithm to assign weights to tags that are not strongly correlated with each other.

While results from lasso and random forest regressions are consistent, their $R^2$ values for each personality trait differ. For example, predictions of conscientiousness scores have the lowest $R^2$ for lasso regression, but the second highest $R^2$ for random forest regression. This difference may be explained by the linearity constraint for lasso regression–if the distribution of personality scores is non-linear, its $R^2$ in lasso regression may be affected.

*4.1.2 Discussion*

Overall, we find that book preferences can potentially be used to predict personality traits:

**Extraversion.** As expected, our findings suggest that extraverts enjoy books with social themes, as described by the tags like `relationships` and `chick lit`. They also seem interested to read about the lives of others, from memoirs to celebrity romance. Curiously, preference for African American literature is also associated with being extraverted. This may be explained by African Americans themselves being more extraverted than white Americans [23]. Since we did not record race in our study, we cannot rule out this explanation. On the other hand, introverts seem to prefer books with themes such as fantasy, science fiction, and supernatural forces, exhibiting a tendency to indulge in imagination. Appreciation of Japanese culture, especially manga and comics, is also associated with introversion. In general, book preferences explain a substantial amount of variation in the extraversion dimension, consistent with the consensus that extraversion is typically a more salient trait to measure.

**Agreeableness.** Our findings suggest that agreeable people enjoy books with family and religious themes, both of which promote positive social relationships. On the flip side, disagreeable individuals seem attracted to dark-themed content such as psychological dramas. Cult classics, known for their controversial narratives, also seem appealing to these individuals who may have fewer qualms about resisting popular opinion. Books with content related to Japan, Italy, and Russia are also read by people who are less agreeable, possibly because people from these cultures tend to score lower on agreeableness compared to Americans [23]. Interestingly, most of the top tags predicting agreeableness are positively correlated with the trait. The absence of consistent tags endorsed by disagreeable people suggests that these people also tend to disagree on what they 'liked'.

**Openness.** Open individuals seem to enjoy intellectually challenging books that the average person may find difficult to complete (e.g. `back-burners`). Their preference for classic literature further reinforces this view, as books of this genre usually take substantial effort to finish. This is consistent with past studies that found openness to be highly correlated with appreciation for art and literature [13]. Our results also show that individuals scoring lower on openness prefer mainstream content that are less cognitively taxing and easier to digest, such as light-fantasy. Content related to Christianity and India are also preferred by readers with low openness, likely due to religious individuals [24] and Indians [23] scoring low on this trait.

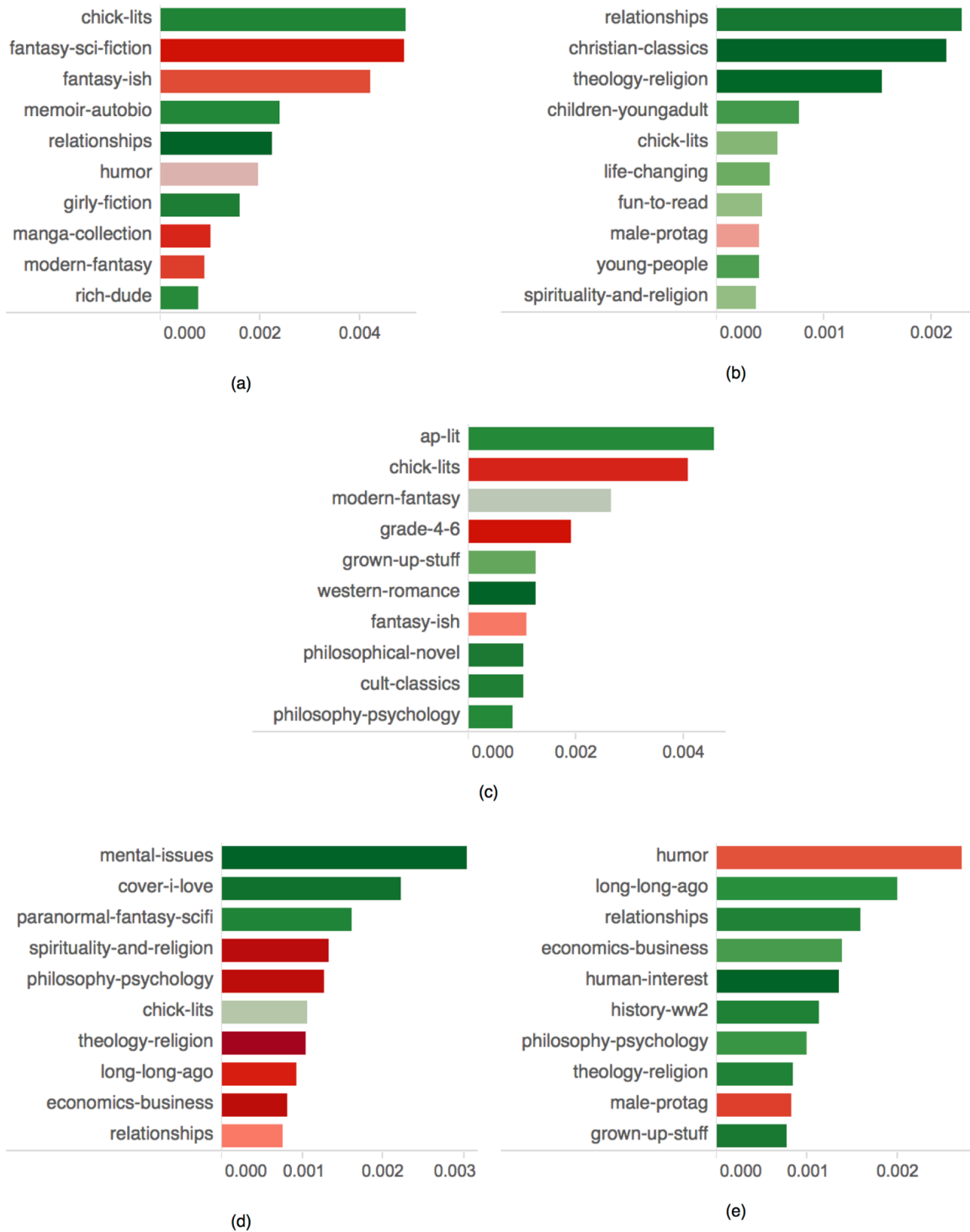

Fig. 2. Top tags that resulted in largest decreases in mean squared error (MSE) in random forest regression analysis for each personality trait. Colors show the correlation between tag and personality trait (red is negative while green is positive). (a) extraversion ($R^2 = 0.35$); (b) agreeableness ($R^2 = 0.24$); (c) openness ($R^2 = 0.47$); (d) neuroticism ($R^2 = 0.31$); (e) conscientiousness ($R^2 = 0.37$).

**Neuroticism.** Neurotic individuals seem to indulge in narratives that reflect their own emotional states, such as

TABLE 3
TOP TAGS IN EXAMPLE CLUSTERS

| Classics | South American | Mystery |
|---|---|---|
| regency-era | south-american | legal-suspense |
| victorian-gothic | paulo | detective-mysteries |
| 2015-classics-challenge | spanish-lit | law-fiction |
| time-s-100-best-novels | translated-literature | fiction-thrillers |
| forced-reads | 21st-century-lit | thriller-detective |
| classic-favs | international-setting | secret-society |

| Philosophy | Girls' Fiction | Paranormal |
|---|---|---|
| philosophical-novel | childrens-realistic-fiction | romance-paranormal-romance |
| spirituality-and-religion | fluffy-fun | wolf-books |
| science-and-technology | mother-daughter-relationships | vamp-stuff |
| philosophy-psychology | roadtrip | uf-or-paranormal |
| arabic-books | women-and-gender | vampire-academy-series |
| politics-economics | grade-4-6 | gods-and-myths |

`sad endings` and `mental issues`. They also appear to enjoy books on alternative realities, in line with the hypothesis that these genres provide a means of escape [4]. Interestingly, neurotic individuals like books with `pretty covers`, possibly due to a gender effect as females tend to score higher in neuroticism than males [25]. On the other hand, emotionally stable individuals prefer self-improvement and other non-fiction content that better reflect reality. In general, we found book preferences to be good predictors of neuroticism, explaining as much as 59% of the variance in this dimension.

**Conscientiousness.** Hardworking people appear to prefer informative content that contributes to their professional development, or that simply boosts their knowledge [6], [15]. On the other hand, people with low conscientiousness scores tend to like lighthearted content (e.g. `humor`) and books aimed at youths (e.g., `teenage-books`). This can be explained by how teenagers tend to score lower than the middle-aged in conscientiousness [26].

In sum, our results show how book preferences can be used to predict one's personality. Besides personality, our findings also reveal cultural differences in book preferences, further supporting the utility of online, user-generated data in deducing more comprehensive profiles of target audiences.

## 4.2 Genre-Level Analysis

Conclusions from our tag-level analyses are based on finer descriptions of book content rather than traditional genres. To test the integrity of tags as book content descriptors, we further group tag clusters into broader genres, which are then used to predict personality scores again.

To obtain genre clusters, we compute the Pearson's correlation between the tf-idf values of books as a proxy for dissimilarity (i.e. distance) between books. Next, the books are clustered using the Partitioning Around Medoids (PAM) algorithm, a form of k-medoid clustering [27]. Like k-means, PAM aims to minimize the distance between cluster members and their respective cluster centers through an iterative algorithm. Unlike k-means however, PAM assigns actual data points as cluster centers. Hence, it is more robust to noise and outliers than k-means because it minimizes the sum of pairwise dissimilarities rather than the sum of squared Euclidean distances.

To determine the optimal number of clusters, we use silhouette width, a measure for data points' similarity within their assigned cluster against their similarity to points in other clusters. For data point $i$, we let $a(i)$ denote the average distance between $i$ and all other data points in the cluster that $i$ is assigned to, and $b(i)$ be the lowest average distance of $i$ to any other cluster. Silhouette $s(i)$ is defined as

$$s(i) = \frac{b(i) - a(i)}{\max\{a(i), b(i)\}} \quad (4)$$

We examined results for 4 to 30 clusters, and eventually chose the 27-cluster solution as it yielded large mean and median silhouette widths across all clusters. These clusters also represented a diverse range of genres that enable comparison with past literature. The composition of each genre in terms of tags, as well as the personality profile of each genre, are presented in the following results section.

### 4.2.1 Results

Genres clusters are given labels that are representative of their member tags. Top tags from example clusters are shown in Table 3. These are the tags that appear most frequently in a genre relative to the entire dataset.

For each genre, we took the median personality scores of all books in that genre cluster, thus generating an overall personality profile for that genre. Fig. 3 shows the aggregated personality profiles for all 27 genres. Size of pie chart slices are normalized to zero mean and unit variance.

A principal component analysis was performed on the aggregated personality scores across genres, and we found that the openness and conscientiousness traits captured the most variation in genre profiles. Thus, for visualization purposes, we plot book genres for these two dimensions in Fig. 4.

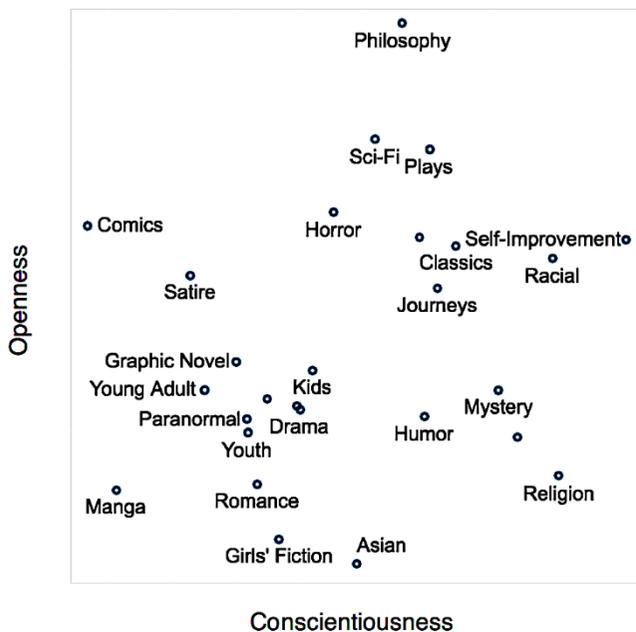

Fig. 4. Genres on the Openness-Conscientiousness dimensions.

### 4.2.2 Discussion

Results from the genre-level analysis are consistent with findings from both existing literature and our tag-level analysis. For instance, people who like *Self-improvement* books are more conscientious than those who like *Comics*, and people who like *Philosophy* books are more open than those who like *Religious* books.

There is one exception, however. While a previous study [6] found that extraverts like horror, our findings suggest the opposite– books with horror themes seem to appeal more to introverts. This discrepancy may be due to the mode of narrative: while our study focuses on books, the other study had examined television shows in addition to books. While horror TV shows may be highly stimulating, the arousal may be muted in books, explaining the lower preference for horror books among our extraverted respondents.

Apart from confirming our earlier results, genre clusters also give new insights. For example, the *Thriller* cluster contains detective and legal elements, which require critical thinking and perhaps even background knowledge on law for a reader to fully appreciate the plot. This may explain why readers who like mystery books also score higher on conscientiousness. Another interesting observation is that the *Classics* cluster has an average profile for all five personality traits. This cluster contains time-honored and household favorites, which would have appealed to most people regardless of personality, thus resulting in a profile that reflects the sample average.

Cultural differences are also apparent. People who like *Asian* books are less open [23], consistent with results from our tag-level analysis. However, people who like *Asian* books are also relatively extraverted, which runs contrary to claims that Asians are more introverted [28]. This discrepancy may be due to Facebook being more attractive to extraverted individuals in the first place [29], thus resulting in a more extraverted Asian user base.

We have shown how personality profiles of readers can be inferred from their preferred content, at both the tag and genre level. A detailed tag-level analysis can provide more resolution on book content, while a broader genre-level analysis can identify associations between tags.

### 4.3 General Reading Disposition

Since personality has been found to correlate with book preferences, it may also correlate with the tendency to like books in the first place. To examine this, we compute correlations between users' personality scores and the number of book pages they 'liked'.

It turns out that correlations are very weak ($r$'s < 0.06) across four of the five traits: conscientiousness, extraversion, neuroticism and, importantly, agreeableness. Although previous studies found that agreeable people tend to evaluate content favorably [10], our study finds a near-zero correlation between agreeableness and the number of books 'liked' on Facebook ($r = -0.02$). A possible explanation may be that while agreeable people are less likely to express dislike to avoid disagreements, they may nonetheless only 'like' a book when they genuinely enjoy the content.

The openness trait, on the other hand, is a relatively strong and significant predictor ($r = 0.12$, $p < 0.001$) of number of books 'liked'. This result lends support to our earlier hypothesis: Open individuals appreciate a wider variety of books, and thus 'like' more books on average.

## 5 LIMITATIONS

We acknowledge that reliance on web data may lead to a few limitations. First, only popular books that had a Facebook page with sufficient 'likes' are included in the analysis. Hence, newer or niche books on the heavy tail of a book popularity distribution may be overlooked. Second, Facebook users have been found to be more extraverted, more narcissistic, and less conscientious than average, and hence they may not be representative of the general population [29]. Third, Facebook 'likes' may be driven by the need for social acceptance or recognition [30], [31], and thus may not be a faithful reflection of a person's preferences.

However, because our findings are in line with existing literature, the above concerns are unlikely to have been significant enough to skew results. In fact, despite sources of noise and idiosyncrasies, user-generated tags have proven to be a rich well of information that not only enabled us to dive deeper into sub-genre preferences, but also to explore broader preference-related behaviors.

## 6 CONCLUSIONS

Findings from our study are consistent across both tag and genre levels of analyses, and also in line with existing literature, thus demonstrating the utility of online user-generated data in profiling target audiences. Besides predicting personality from book preferences, user tags allow us to uncover unexpected insights, such as cultural differences, book reading behaviors (e.g. 'back-burner'), and other non-content factors affecting preferences (e.g. 'pretty covers').

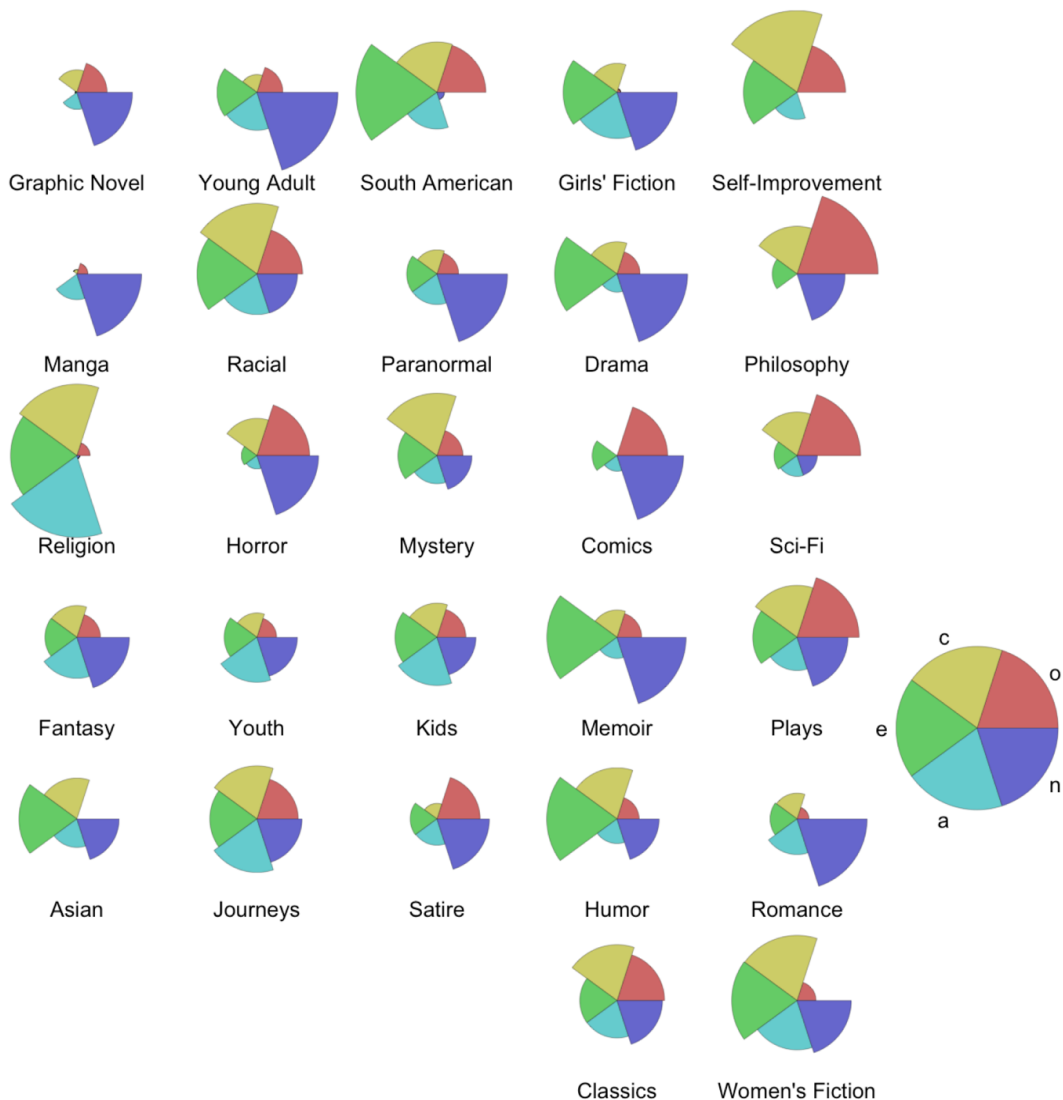

Fig. 3. Personality profiles of genres.

Future research can incorporate additional dimensions such as year of publication, which may allow us to track the evolution of genres. For instance, vanilla love stories in the romance genre seem to be increasingly overtaken by vampire-related themes, with series such as *Twilight (2005-2008)* and *Vampire Academy (2007-2010)*. Trends like these may be overlooked if books are analyzed by genre instead of tags. Another possible avenue of research may be to examine popular combinations of tags within books. Findings may help authors identify unique tag combinations to spin fresh story plots.

With growing online activity, we believe that large, user-generated datasets, as well as the ability to parse them effectively, can play an important role in the study of arts and social sciences fields, such as literature, psychology, and marketing.

## ACKNOWLEDGMENT

The authors thank Dr. Michal Kosinski for his valuable feedback, and Goodreads.com for allowing us to crawl for data.

## REFERENCES

[1] M. Djikic, K. Oatley, S. Zoeterman, and J. B. Peterson, "On being moved by art: How reading fiction transforms the self," Creativity Research Journal, vol. 21, no. 1, pp. 24–29, Feb. 2009.
[2] R. A. Mar, K. Oatley, and J. B. Peterson, "Exploring the link between reading fiction and empathy: Ruling out individual differences and examining outcomes," Communications, vol. 34, no. 4, Jan. 2009.
[3] W. C. Tirre and S. Dixit, "Reading interests: Their dimensionality and correlation with personality and cognitive factors," Personality and Individual Differences, vol. 18, no. 6, pp. 731–738, Jun. 1995.
[4] G. Kraaykamp and K. van Eijck, "Personality, media preferences, and cultural participation," Personality and Individual Differences, vol. 38, no. 7, pp. 1675–1688, May 2005.
[5] I. Cantador, I. Fernández-Tobías, and A. Bellogín, "Relating personality types with user preferences in multiple entertainment domains," CEUR Workshop Proceedings, vol. 997, 2013.

[6] P. Rentfrow, L. Goldberg, and R. Zilca, "Listening, watching, and reading: The structure and correlates of entertainment preferences," Journal of personality., vol. 79, no. 2, pp. 223–58, Jul. 2010.
[7] M. Galesic and M. Bosnjak, "Effects of questionnaire length on participation and indicators of response quality in a web survey," Public Opinion Quarterly, vol. 73, no. 2, pp. 349–360, Jan. 2009.
[8] M. Kosinski, D. Stillwell, T. Graepel, "Private traits and attributes are predictable from digital records of human behavior," Proceedings of the National Academy of Sciences, vol. 110, no. 15, pp. 5802–5805, Sep. 2013.
[9] P. T. Costa, R. R. McCrae, "The revised neo personality inventory (NEO-PI-R)," The SAGE Handbook of Personality Theory and Assessment, vol. 2, pp. 179–198, 2008.
[10] J. B. Weaver, H.-B. Brosius, and N. Mundorf, "Personality and movie preferences: A comparison of American and German audiences," Personality and Individual Differences, vol. 14, no. 2, pp. 307–315, Feb. 1993.
[11] M. T. Soto-Sanfiel, L. Aymerich Franch, and E. Romero, "Personality in interaction: How the big Five relate to the reception of interactive narratives," Comunicación y sociedad = Communication & Society, vol. 27, no. 3, pp. 151–186, 2014.
[12] A. Furnham and J. Walker, "Personality and judgements of abstract, pop art, and representational paintings," European Journal of Personality, vol. 15, no. 1, pp. 57–72, Jan. 2001.
[13] I. C. McManus and A. Furnham, "Aesthetic activities and aesthetic attitudes: Influences of education, background and personality on interest and involvement in the arts," British Journal of Psychology, vol. 97, no. 4, pp. 555–587, Nov. 2006.
[14] J. C. Conway and A. M. Rubin, "Psychological predictors of television viewing motivation," Communication Research, vol. 18, no. 4, pp. 443–463, Aug. 1991.
[15] A. S. Gerber, G. A. Huber, D. Doherty, and C. M. Dowling, "Personality traits and the consumption of political information," American Politics Research, vol. 39, no. 1, pp. 32–84, Sep. 2010.
[16] L. R. Goldberg et al., "The international personality item pool and the future of public-domain personality measures," Journal of Research in Personality, vol. 40, no. 1, pp. 84–96, Feb. 2006.
[17] S. C. Rife, K. L. Cate, M. Kosinski, and D. Stillwell, "Participant recruitment and data collection through Facebook: The role of personality factors," International Journal of Social Research Methodology, pp. 1–15, Sep. 2014.
[18] M. Kosinski, Y. Bachrach, P. Kohli, D. Stillwell, and T. Graepel, "Manifestations of user personality in website choice and behaviour on online social networks," Machine Learning, vol. 95, no. 3, pp. 357–380, Jan. 2014.
[19] G. Park et al., "Automatic personality assessment through social media language," Journal of Personality and Social Psychology, vol. 108, no. 6, pp. 934–952, 2015.
[20] W. Youyou, M. Kosinski, and D. Stillwell, "Computer-based personality judgments are more accurate than those made by humans," Proceedings of the National Academy of Sciences, vol. 112, no. 4, pp. 1036–1040, Jan. 2015.
[21] J. D. Choi and M. Palmer, "Fast and robust part-of-speech tagging using dynamic model selection," Association for Computational Linguistics, 2012, pp. 363–367.
[22] M. Ankerst, M. M. Breunig, H.-P. Kriegel, and J. Sander, "OPTICS: ordering points to identify the clustering structure," ACM SIGMOD Record, vol. 28, no. 2, pp. 49–60, Jan. 1999.
[23] J. Allik and R. R. McCrae, "Toward a geography of personality traits: Patterns of profiles across 36 cultures," Journal of Cross-Cultural Psychology, vol. 35, no. 1, pp. 13–28, Jan. 2004.
[24] V. Saroglou, "Religion and the five factors of personality: A meta-analytic review," Personality and Individual Differences, vol. 32, no. 1, pp. 15–25, Jan. 2002.
[25] D. P. Schmitt, A. Realo, M. Voracek, and J. Allik, "Why can't a man be more like a woman? Sex differences in big Five personality traits across 55 cultures," Journal of Personality and Social Psychology, vol. 94, no. 1, pp. 168–182, 2008.
[26] M. B. Donnellan and R. E. Lucas, "Age differences in the big five across the life span: Evidence from two national samples," Psychology and Aging, vol. 23, no. 3, pp. 558–566, 2008.
[27] L. Kaufman and P. J. Rousseeuw, Finding Groups in Data: An Introduction to Cluster Analysis (Wiley Series in Probability and Statistics). John Wiley & Sons, 2008, ch. 2, pp. 68–125.
[28] D. P. Schmitt, J. Allik, R. R. McCrae, and V. Benet-Martinez, "The geographic distribution of big Five personality traits: Patterns and profiles of human self-description across 56 nations," Journal of Cross-Cultural Psychology, vol. 38, no. 2, pp. 173–212, Mar. 2007.
[29] T. Ryan and S. Xenos, "Who uses Facebook? An investigation into the relationship between the big Five, shyness, narcissism, loneliness, and Facebook usage," Computers in Human Behavior, vol. 27, no. 5, pp. 1658–1664, Sep. 2011.
[30] H. Gangadharbatla, "Facebook me: Collective self-esteem, need to belong, and Internet self-efficacy as predictors of the igeneration's attitudes toward social networking sites," Journal of Interactive Advertising, vol. 8, no. 2, pp. 5–15, Mar. 2008.
[31] L. E. Buffardi and W. K. Campbell, "Narcissism and social networking web sites," Personality and Social Psychology Bulletin, vol. 34, no. 10, pp. 1303–1314, Jul. 2008.

**Ng Annalyn** was a research associate with Disney Research Pittsburgh and is currently employed by Singapore's Ministry of Defence. She received her M.Phil. degree in Psychology with the University of Cambridge, where she mined consumer data for targeted advertising and programmed cognitive tests for job recruitment with the Cambridge Psychometrics Centre. She has a B.Sc. in Psychology and Economics from the University of Michigan (Ann Arbor), where she was also an undergraduate statistics tutor. Her research interests include machine learning applications in social sciences. She is the author of the book: "*Numsense! Data Science for the Layman*".

**Maarten W. Bos** received his MS degree in Social Psychology from the University of Amsterdam, and his PhD degree from the Radboud University in The Netherlands. After a postdoctoral fellowship at the Harvard Business School, he is currently a Research Scientist at Disney Research. His research interests include decision science and behavioral economics. He has high impact publications in decision science, and he is a member of the Society for Personality and Social Psychology, the Association for Psychological Science, and the Society for Judgment and Decision Making.

**Leonid Sigal** is a Senior Research Scientist at Disney Research Pittsburgh and an adjunct faculty at Carnegie Mellon University. Prior to this he was a postdoctoral fellow in the Department of Computer Science at University of Toronto. He completed his Ph.D. at Brown University in 2008; he received his B.Sc. degrees in Computer Science and Mathematics from Boston University (1999), his M.A. from Boston University (1999), and his M.S. from Brown University (2003). From 1999 to 2001, he worked as a senior vision engineer at Cognex Corporation, where he developed industrial vision applications for pattern analysis and verification. Leonid's research interests mainly lie in the areas of computer vision, machine learning, and computer graphics. He has published more than 50 peer reviewed papers in venues and journals in in these fields (including publications in PAMI, IJCV, CVPR, ICCV, ECCV, NIPS, UAI, and ACM SIGGRAPH). His work received the Best Paper Awards at the AMDO conference in 2006 / 2012 and at WACV in 2014. He has also coedited the book Guide to Visual Analytics of Humans: Looking at People (Springer, 2011).

**Boyang "Albert" Li** is a Research Scientist at Disney Research, where he directs the Narrative Intelligence group. He obtained his Ph.D. in Computer Science from Georgia Institute of Technology in 2014, and his B. Eng. from Nanyang Technological University, Singapore in 2008. His research interests include computational narrative intelligence, or the creation of Artificial Intelligence that can understand, craft, tell, direct, and respond appropriately to narratives, and understanding how human cognition comprehends narratives and produces narrative-related affects. He has authored and co-authored more than 30 peer-reviewed papers in international journals and conferences.